# Structural, Ferroelectric, Magnetic and Magnetoelectric Response in Multiferroic (1-x)Bi(Ni$_{1/2}$Ti$_{1/2}$)O$_3$-PbTiO$_3$/xNi$_{0.6}$Zn$_{0.4}$Fe$_2$O$_4$ Particulate Composites

*Rishikesh Pandey*[*,1,2], *Uma Shankar,*[2] *Sher Singh Meena*[3] *and Akhilesh Kumar Singh*[2]

[1]*Motihari College of Engineering, Motihari-845401*

[2]*School of Materials Science & Technology, Indian Institute of Technology (Banaras Hindu University) Varanasi- 221005, India*

[3]*Solid State Physics Division, Bhabha Atomic Research Centre, Mumbai-400085, India*

**Abstract**

Multiferroic particulate composites have been fabricated by taking the morphotropic phase boundary composition of ferroelectric phase Bi(Ni$_{1/2}$Ti$_{1/2}$)O$_3$-PbTiO$_3$ and magnetic phase (Ni,Zn)Fe$_2$O$_4$. The ferroelectric phase has coexisting monoclinic and tetragonal perovskite structures with space group *Pm* and *P4mm*, respectively whereas the magnetic phase has spinel cubic structure with space group *Fd$\bar{3}$m*. Rietveld structural analysis for the each components of composite reveals that the tetragonality (c/a) of the ferroelectric phase continuously increases with increasing the concentration of magnetic phase suggesting partial ionic diffusion between ferroelectric and magnetic phases, after composite formation. Composition dependent Mössbauer spectra of (1-x)Bi(Ni$_{1/2}$Ti$_{1/2}$)O$_3$-PbTiO$_3$/x(Ni,Zn)Fe$_2$O$_4$ reveals the superparamagnetic like behavior for the ferroelectric rich composition with x=0.2. The magnetic ordering increases for the composition with x=0.4 and 0.6 which completely transform into ferrimagnetic for the composition with x=0.9 for the magnetic phase rich compositions. Unlike the ferroelectric or magnetic components which do not exhibit the magnetoelectric response separately, large value of magnetoelectric coefficient (> 30 mV/Oe-cm) in (1-x)Bi(Ni$_{1/2}$Ti$_{1/2}$)O$_3$-PbTiO$_3$/xNi$_{0.6}$Zn$_{0.4}$Fe$_2$O$_4$ composite makes it promise for multifunctional applications.

\* Corresponding author
  E-mail: rishikeshbhu09@gmail.com

# 1. Introduction

In recent years, multiferroics have attracted extensive research interest due to the coupled ferroelectric and magnetic ordering termed as magnetoelectric (ME) coupling. Magnetoelectric coupling in multiferroic materials provides wide opportunities for designing the novel devices such as and sensors, transducers, actuators, spinotronics, data storage etc [1-3]. However due to complicated structural chemistry, very few multiferroics are synthesized or found in the nature. Presence of the vacant d-orbit ($d^0$) at the B-site atom in ferroelectric phase and unpaired electrons in d-orbit ($d^n$) of magnetic phase are the contradictory requirement for the coexistence of multiferroic systems [4-5]. Most of the multiferroics synthesized in the form of single phase till date exhibit very small ME coefficient which respond at low temperature and found to be not suitable for device application [6-9]. This drawback of single phase multiferroics forced researchers to investigate some new materials which exhibit high ME response at ambient temperature. Multiferroic composites came into the consideration to overcome the drawback of single phase multiferroics. Recently, multiferroic composites have been attracted great interest due to the coupled ferroelectric and magnetic ordering and high ME coefficient well above the room temperature [9-10]. Giant magnetoelectric coefficient (> 1 V/cm-Oe) has been reported by Dong et al. in the bulk composite of Terfenol-D/Pb($Mg_{1/3}Nb_{2/3}$)$O_3$-$PbTiO_3$ [11]. Magnetoelectric coupling in multiferroic composites is a multiplicative property of the ferroelectric and magnetic phases mechanically mediated by strain [2]. On the application of the magnetic field, magnetic phase changes the shape due to magnetostriction. The developed strain in the magnetic phase passed in the piezoelectric phase through the mechanical contact between two phases, which results in polarization in the piezoelectric phase. Similarly, strain is developed in the piezoelectric phase on the application of electric field which is transferred mechanically to the magnetic phase. Magnetostrictive magnetic phase changes the shape which results into the magnetization in the magnetic phase [12,13]. The direct and converse ME-effect in the multiferroic composites have been defined by equations (i) and (ii) as [12,13]

$$\text{Direct ME effect} = \frac{Magnetic}{Mechanical} \times \frac{Mechanical}{Electrical} \quad \ldots\ldots\ldots\ldots \quad (i)$$

$$\text{Converse ME effect} = \frac{Electrical}{Mechanical} \times \frac{Mechanical}{Magnetic} \quad \ldots\ldots\ldots \quad (ii)$$

Thus, a ferroelectric system with large tetragonality is required for the best coupling between ferroelectric and magnetic phases. Following this scheme, several multiferroic particulate composites have been designed by taking the various combinations of ferroelectric and magnetic phases such as $BaTiO_3/CoFe_2O_4$ [10,14], $PbZr_xTi_{1-x}O_3/(Ni,Zn)Fe_2O_4$ [15,16], $(Ba,Ca)(Zr,Ti)O_3/CoFe_2O_4$ [17,18], $BiFeO_3/Ni_{0.5}Zn_{0.5}Fe_2O_4$ [19] etc. In the particulate composites, ME-coupling takes place at the interface of the grains during sintering therefore, proper mechanical contact in between the grains, high piezoelectric/magnetostrictive coefficients and high resistivity of ferroelectric as well as magnetic phases are the essential requirement for the formation of high performance particulate composite [15,20,21].

In the present work, we have fabricated a multiferroic particulate composite by taking the ferroelectric phase $0.51Bi(Ni_{1/2}Ti_{1/2})O_3$-$0.49PbTiO_3$ (BNT-PT) and the magnetic phase $Ni_{0.6}Zn_{0.4}Fe_2O_4$ (NZFO) as $(1-x)0.51Bi(Ni_{1/2}Ti_{1/2})O_3$-$0.49PbTiO_3/xNi_{0.6}Zn_{0.4}Fe_2O_4$ ((1-x)BNT-PT/xNZFO). 'Bi/Pb'-based ferroelectric solid solutions with of type $Bi(M'_{1/2}M''_{1/2})O_3$-$PbTiO_3$ have been found to be of great interest due to its high $T_c$, large $d_{33}$, good electromechanical response and relaxor nature. Sintering temperature for the various $Bi(M'_{1/2}M''_{1/2})O_3$-$PbTiO_3$ type solid solutions is quite lower (~ 1000 $^oC$) than that of widely used classical ferroelectrics like $BaTiO_3$ and $PbZr_xTi_{1-x}O_3$ (~ 1300 $^oC$), therefore its composites can also be synthesized at low temperatures as compared to them [10,14-16]. It's one of the component $Bi(M'_{1/2}M''_{1/2})O_3$ can accumulate various types of divalent ($M'$) and tetravalent ($M''$) cations which frequently forms the ferroelectric solid solutions with $PbTiO_3$ such as $Bi(Ni_{1/2}Ti_{1/2})O_3$–$PbTiO_3$ (BNT-PT), $Bi(Mg_{1/2}Ti_{1/2})O_3$–$PbTiO_3$ (BMT-PT), $Bi(Mg_{1/2}Zr_{1/2})O_3$–$PbTiO_3$ (BMZ-PT), $Bi(Zn_{1/2}Ti_{1/2})O_3$–$PbTiO_3$ (BZT-PT), $Bi(Co_{1/2}Ti_{1/2})O_3$–$PbTiO_3$, $Bi(Zn_{1/2}Zr_{1/2})O_3$–$PbTiO_3$ and $Bi(Zn_{1/2}Sn_{1/2})O_3$–$PbTiO_3$ [22-27]. Few of them exhibit morphotropic phase boundary (MPB) which is the coexistence region of two ferroelectric phases where the piezoelectric response gets maximized.

The structure of BNT-PT taken in this study has coexisting tetragonal (*P4mm*) and monoclinic (*Pm*) structures [22]. BNT-PT has been reported to exhibit large longitudinal piezoelectric coefficient ($d_{33}$ ~ 260 pC/N) and high paraelectric to ferroelectric phase transition temperature ($T_c$ ~ 400 $^0C$). Due to presence of $Ni^{2+}$ ions, BNT-PT also exhibits weak ferromagnetic nature for MPB compositions. However, it has been proposed that higher concentration of BNT weaken the superexchange interaction which diminishes the ferromagnetic ordering [23-24]. NZFO has spinel cubic structure with space group $Fd\bar{3}m$. The Curie temperature of NZFO taken in the present work is ~ 350 $^0C$, below which it shows ferrimagnetic nature [28]. Therefore, it is possible to design a multiferroic composite exhibiting magnetoelectric response well above the room temperature taking piezoelectric

BNT-PT and magnetostrictive NZFO components. Our structural analysis reveals that the tetragonality of ferroelectric phase BNT-PT increases with increasing the fraction of magnetic phase NZFO. Large tetragonality will produce the significant strain and thus strong coupling between two phases.

## 2. Experimental details

(1-x)BNT-PT/xNZFO particulate composites were synthesized by conventional solid-state route. To prepare $Ni_{0.6}Zn_{0.4}Fe_2O_4$, stoichiometric amounts of NiO (Qualigence, 99.9%), ZnO (Qualigence, 99%) and $Fe_2O_3$ (Himedia, 99%) chemicals were mixed in ball milling containing zirconia jars and zirconia balls for 6 h with acetone as mixing media. The powder mixture was calcined at 800 $^0$C for 6 h. Similarly, BNT-PT was also prepared by conventional solid state route using $Bi_2O_3$ (Himedia, 99%), PbO (Qualigence, 98%), NiO (Qualigence, 99.9%) and $TiO_2$ (Himedia, 99%) chemicals. Details of sample preparation for BNT-PT have been reported, elsewhere [22]. Different compositions of (1-x)BNT-PT/xNZFO composites containing $Ni_{0.6}Zn_{0.4}Fe_2O_4$ as a magnetic phase and $0.51Bi(Ni_{1/2}Ti_{1/2})O_3$-$0.49PbTiO_3$ as a ferroelectric phase were prepared by solid state ceramic route. In this process, stoichiometric amount of calcined powders of BNT-PT and NZFO were first ball-milled with acetone in zirconia jars (containing zirconia balls) for 6 h and then dried. 2% polyvinyl alcohol (PVA) solution in water used as binder was mixed with the powders. The powder mixture was pressed into the form of pellets of diameter ~ 12 mm and thickness ~ 1.5 mm using a stainless-steel die and uniaxial hydraulic press at an optimized load of 65 kN. Before sintering, the green pellets were kept at 500 $^0$C for 10 h to burn out the binder. The pellets were sintered at 1000 $^0$C for 3 h. Powder X-ray diffraction (XRD) pattern was recorded using an 18 kW rotating anode Cu-target based RIGAKU (Japan) powder X-ray diffractometer fitted with a graphite monochromator in the path of diffracted beam. For (*P-E*) measurement, fired on silver paste was subsequently applied on both flat surface of the pellet. It was first dried around 120 $^0$C in an oven and then cured by firing at 500 $^0$C for 5 minutes. (*P-E*) hysteresis measurement was carried out by Radiant (P-E) loop tracer (USA). The microstructural and energy dispersive spectroscopy (EDS) studies were done by ZEISS, Supra-40 scanning electron microscope (SEM). Before the microstructural study, sintered pellets were sputter coated with Pd/Au alloys. FULLPROF program [29] was used for Rietveld refinement of the structure of different compositions of composite. Pseudo-Voigt function was used to define the peak profiles and sixth-order polynomial was used to fit the background. The magnetic phase NZFO was refined by spinel cubic structure with space group $Fd\bar{3}m$ (Space group # 227). The ferroelectric phase BNT-PT was fitted by tetragonal structure with $P4mm$ space group (Space group # 99). The Mössbauer spectra of BNT-PT/NZFO composites were recorded using a Mössbauer spectrometer operated in constant acceleration mode (triangular wave) in transmission geometry at room temperature. The source employed was Co-57 in Rh matrix of strength 50 mCi. The calibration of the velocity scale was done by using an enriched Fe metal foil. All Mössbauer spectra were fitted by using a Win-Normos fit program. To measure the ME coefficient, the sintered pellets were poled both electrically and magnetically. In electric poling, samples were heated at 100 $^0$C in the presence of applied external electric field ~ 2.0 kV/mm and subsequently cooled to room temperature in the presence of applied electric field. After the electric poling, the samples were poled magnetically at room temperature by applying an external DC-magnetic field of 6 kOe. The dynamic ME response was obtained by changing the DC magnetic field and then measuring the output signal in the form of voltage at applied small AC magnetic field of 3 Oe.

## 3. Results and discussion

### 3.1. Crystal structure of composite components

Fig.1 shows the powder XRD profile of (1-x)BNT-PT/xNZFO composites for the compositions with x=0 to 1.0 at an equal compositional interval of 0.1. The XRD peaks corresponding to BNT-PT and NZFO are marked by 'b' and 'n', respectively. As could be seen in Fig.1, no impurity phases have formed during composite formation and all the peaks are appearing due to magnetic NZFO and ferroelectric phases BNT-PT only. As the fraction of NZFO phase is increased in the composite, the intensity of XRD profiles corresponding to cubic NZFO increases while the intensity of the XRD profiles corresponding to tetragonal perovskite BNT-PT decreases. All the XRD peaks in Fig.1 are indexed with tetragonal ($P4mm$) structure for BNT-PT and spinel cubic ($Fd\bar{3}m$) structure for NZFO. We carried out Rietveld structural refinement for all the compositions of composite to check the phase purity and determine structural parameters precisely. Rietveld structural refinement for NZFO using cubic space group $Fd\bar{3}m$ is shown in Fig.2(a). In the cubic spinel structure of NZFO, $Zn^{2+}$/$Fe_I^{3+}$ ions occupy the tetrahedral sites 8(a) at (1/8, 1/8, 1/8), $Ni^{2+}$/$Fe_{II}^{3+}$ ions occupy the octahedral sites 16(d) at (1/2, 1/2, 1/2) and $O^{2-}$ ions occupy the 32(e) sites at (x+1/4, x+1/4, x). The refined structural parameters and agreements factors are listed in Table 1 for NZFO. Refined structural parameters and agreements factors of BNT-PT are reported elsewhere [22]. Fig.2(b) shows the Rietveld structural refinement for the composite composition with x=0.5 using tetragonal space group *P4mm* for the ferroelectric phase BNT-PT and cubic space group $Fd\bar{3}m$ for the magnetic phase NZFO. Very good fit between the observed and calculated XRD profiles confirms the formation of ideal composite without any additional unwanted impurity phases. The refined structural parameters and agreements factors are listed in Table 2 for BNT-PT/NZFO composite (x=0.5).

A rigorous structural analysis of the ferroelectric phase BNT-PT for the different compositions of the composite has been shown in Fig.3. As shown in Fig.3, (200) XRD peak shift towards the higher 2θ side depicts the diminishing nature of 'a' lattice parameter. Just opposite, (002) peaks shift along the lower 2θ side reveals the increasing trend of 'c' lattice parameter with increasing the content of NZFO in the composite for the compositions x=0-0.8. However, for the composition with x=0.9, both (200) and (002) peaks disappeared and a new peak (024) corresponding to rhombohedral ($R\bar{3}c$) structure appeared in between them. Rietveld fit for the ferroelectric phase of the composition with x=0.9 using rhombohedral $R\bar{3}c$ space group is shown in Fig.4(a). The variation of lattice parameters and tetragonality of ferroelectric phase BNT-PT with composite composition in the composition range x=0-0.8 are shown in Fig.4(b). The 'c' lattice parameter of the ferroelectric phase BNT-PT increases while 'a' lattice parameter decreases with increasing NZFO concentration upto x=0.8. The lattice parameter of NZFO (Fig.4(c)) reveals diminishing nature with increasing the fraction of BNT-PT in the composite suggesting the partial ionic diffusion between ferroelectric phase BNT-PT and magnetic phase NZFO which modify the lattice parameters after the composite formation. Fig.4(b) depicts the drastic enhancement in the tetragonality (c/a) of BNT-PT by increasing the content of NZFO in the composite. The tetragonality of pure BNT-PT is 1.02 [22] which is much lower than the tetragonality 1.06 of $PbTiO_3$ [30]. The tetragonality of BNT-PT increases upto 1.11 for x=0.8 which is significantly higher than the tetragonality of $PbTiO_3$. On further increasing the fraction of NZFO (for the composition x=0.9), the structure of BNT-PT changes in to rhombohedral phase. Rietveld structural profile fitting with XRD peaks marked by rhombohedral space group $R\bar{3}c$ as shown in Fig.4(a) depicts clear splitting in the (111) pseudocubic reflection (2θ ~ 39°) suggests that the structure is rhombohedral which is singlet in the case of tetragonal phase. Superlattice reflection (113) has been marked by asterisk in Fig.4(a). The lattice parameters for the ferroelectric phase having rhombohedral structure (x=0.9) are obtained to be $a_H$=5.5759(2) Å and $c_H$=13.7585(2) Å in the hexagonal setting. The spontaneous lattice strain (*s*) for rhombohedral ($R\bar{3}c$) phase of x=0.9 is calculated by equation (iii) [31,38]

$$s = \frac{c_H}{\sqrt{6}\, a_H} - 1 \quad \text{……………………..} \quad (iii)$$

which comes out to be ~ 0.007 is ~ 16 times lower than that of x=0.8 composition. Only few studies on the structural evaluation of the components of composite have been reported in the literatures. In Pb(ZrTi)O$_3$/Ni$_{1-x}$Zn$_x$Fe$_2$O$_4$ (PZT/NZFO) and Pb(ZrTi)O$_3$/Co$_{1-x}$Zn$_x$Fe$_2$O$_4$ (PZT/CZFO) composites, Srinivasan et. al. [32] have reported that the lattice parameter of NZFO and CFZO increases with increasing the concentration of Zn. These authors have reported that the tetragonality of PZT decreases in multilayer PZT/NZFO and PZT/CZFO composites but structural modification is not significant in PZT for bulk PZT/NZFO composite with the increasing concentration of Zn. It has been also reported [16] that the tetragonality of PZT decreases and finally transforms into cubic structure in bulk PZT/NZFO composite with increasing NZFO. In CoFe$_2$O$_4$/BaTiO$_3$ composite, both tetragonality of BaTO$_3$ and lattice parameter of CoFe$_2$O$_4$ enhance with increasing the content of CoFe$_2$O$_4$ [33]. As magnetoelectric coupling is highly influenced by crystallographic and magnetic symmetry, slight change in the crystal structure may drastically modify the magnetoelectric coupling coefficient as reported in BaTiO$_3$/La$_{0.67}$Sr$_{0.33}$MnO$_3$ epitaxial film [2,34]. In our case, unusual increase in the tetragonality of ferroelectric phase occurs due to partial ionic diffusion between the components of composites [35]. Earlier authors have reported [36] the enhancement in the tetragonality of (1-x)Bi(Zn$_{1/2}$Ti$_{1/2}$)O$_3$-xPbTiO$_3$ solid solution with increasing Bi(Zn$_{1/2}$Ti$_{1/2}$)O$_3$ concentration. Similarly, enhancement in the tetragonality of (1-x)BiFeO$_3$-xPbTiO$_3$ solid solution has been also observed with the increasing BiFeO$_3$ concentration which finally turns in to rhombohedral ($R\bar{3}c$) phase for higher concentration of BiFeO$_3$ [37-38]. Based on XRD patterns shown in Fig.3 and reported results by earlier authors [36-37], it seems that diffusion of Fe$^{3+}$ or Zn$^{2+}$ ions is taking place from NZFO to B-site of ferroelectric BNT-PT phase which results in enhancement of tetragonality of BNT-PT.

*3.2. Microstructure and elemental analysis*

Evidence of ionic diffusion has been examined by SEM and EDS characterizations of (1-x)BNT-PT/xNZFO composites. SEM image for the compositions with x=0.3 and 0.7 are shown in Figs.5 (a) and (b). Two types of grains morphology (dark and white) corresponding to magnetic NZFO and ferroelectric BNT-PT can be visibly distinguished from the SEM image. We carried out EDS analysis to identify the grains of each phase present in the composite. EDS spectrum of the dark grains for the composition x=0.3 (Fig.5(c)) reveals the presence of Ni$^{2+}$, Zn$^{2+}$ and Fe$^{3+}$ ions which signifies to the grains corresponding to magnetic

NZFO phase. EDS spectrum of the white grains (Fig.5(d)) for the composition x=0.3 reveals the presence of $Bi^{3+}$, $Pb^{2+}$, $Ti^{4+}$ and $Ni^{2+}$ ions which signifies the grains of ferroelectric BNT-PT phase. Similarly, EDS spectrum of the dark grains for the composition x=0.7 (Fig.5(e)) corresponds to the magnetic NZFO phase and white grains (Fig.5(f)) signify the ferroelectric phase BNT-PT. It is interesting to notice that EDS spectrum for NZFO phase shows the peak of $Ti^{4+}$ ions also marked by arrow in Fig5(c) and Fig.5(e). Meanwhile, EDS spectrum for BNT-PT depicts the peak of $Fe^{3+}$ ion marked by arrow in Fig5(d) and Fig.5(f). These peaks are observable in the EDS spectrum of both compositions with x=0.3 and 0.7 which reveal the diffusion of $Ti^{4+}$ ions from ferroelectric phase (BNT-PT) to magnetic phase (NZFO) and $Fe^{3+}$ ions from magnetic phase to ferroelectric phase (BNT-PT) in (1-x)BNT-PT/xNZFO composite. SEM and EDS characterization depict the diffusion of only $Fe^{3+}$ ions (not $Zn^{2+}$ ions) in ferroelectric phase of the composite. Zhang et al [35] have also reported the significant grain size variations due to partial ionic diffusion in $0.94Bi_{0.5}Na_{0.5}TiO_3$-$0.06BaTiO_3$/ZnO composite. Narayan et al. [38] have reported that unusually large strain changes the ground state structure of the ferroelectric phase of $BiFeO_3$-$PbTiO_3$ from tetragonal ($P4mm$) to rhombohedral ($R\bar{3}c$) to minimize the ground state energy of system. This may be the possible reason for the tetragonal to rhombohedral transformation in (1-x)BNT-PT/xNZFO for x=0.9 which shows the lattice strain ~ 0.7 % even lower than $PbTiO_3$ (6 %) and x=0.8 (~ 11 %). However, such structural transformation has not been reported in $Bi(Zn_{1/2}Ti_{1/2})O_3$-$PbTiO_3$ [27]. Distribution of grain size of ferroelectric BNT-PT and magnetic NZFO for x=0.3 and 0.7 has been shown with histogram by Lorentzian curve fitting in Fig.6. It is evident from Fig.6(a) that the most of grains lies in the range of (2 - 5) micron for the ferroelectric BNT-PT with x=0.3. However, few larger grains > 6 microns are also present for BNT-PT for x=0.3. Fig.6(b) shows that most of grains lies in the range of (0.5 - 3) microns for NZFO with x=0.3. Variation of grain size for BNT-PT with x=0.7 has been shown in Fig.6(c). Most of the grains lies in the range of (0.5 - 3) microns for BNT-PT with x=0.7. Distribution of grain size for NZFO with x=0.7 has been shown in Fig.6(d). Large numbers of grains were found below 4 microns for NZFO with x=0.7. However, few grains of size > 5 microns were also obtained for NZFO with x=0.7. It is obvious that average grain size of ferroelectric phase BNT-PT going to decrease with increasing the concentration of NZFO in composite.

*3.3. Ferroelectric response*

Fig.7 shows the (*P-E*) hysteresis loop for the different compositions of (1-x)BNT-PT/xNZFO composites with x=0, 0.1, 0.3, 0.5, 0.7 and 0.9. All the compositions of (1-x)BNT-PT/xNZFO composite exhibit ferroelectric (*P-E*) hysteresis loop. As expected, the saturation/remanent polarization decrease with increasing the NZFO concentration in the composite. The decrease in the saturation/remanent polarization is due to the reduced phase fraction of ferroelectric phase in the composite. Uniform distribution of NZFO in BNT-PT matrix clamps the reorientation of ferroelectric domains in the direction of applied electric field and thus reduces the saturation polarization. Lowering of polarization has been also reported in bulk $Ba(Zr_{0.2}Ti_{0.8})O_3$-$(Ba_{0.7}Ca_{0.3})TiO_3$/$CoFe_2O_4$ composite [17].

*3.4. Magnetic and magnetoelectric characterizations*

The Mössbauer spectra of (1-x)BNT-PT/xNZFO composite for the compositions with x=0.2, 0.4, 0.6 and 0.9 recorded at room temperature are shown in Fig.7. The Mössbauer spectra of compositions with x=0.2, 0.4, 0.6 shows three sextets and a doublet whereas for composition x=0.9 shows four sextets. The analysis results are given in Table 3. The assignments of the sextets have been done based on the Isomers shift ($\delta$) and hyperfine filed values. The value of hyperfine filed for octahedral site in $Ni_{0.6}Zn_{0.4}Fe_2O_4$ sample is lower than that of tetrahedral site [39]. The value of $\delta$ is higher than that of tetrahedral site in all spinel ferrites [40-42]. The range of values of isomer shift (0.202 - 0.599 mm/s) indicates that iron exists in $Fe^{3+}$ valence state with high spin configuration in the all samples [51], no traces of $Fe^{2+}$ ions has been confirmed by Mossbauer spectroscopy [43-47]. The broad sextets C and D (for composition x=0.9) may be associated with the range of nanoparticles in the sample. The week sextet pattern in this sample indicates the presence of a small fraction of magnetic ordered particles and the intense doublet pattern for x=0.2-0.6 of (1-x)BNT-PT/xNZFO composites indicate that the majority of particles are ultrafine particles with superparamagnetic behavior [48-52].

The ME response of (1-x)BNT-PT/xNZFO composites for the compositions with x=0.2, 0.3, 0.5, 0.8 and 0.9 are shown in Fig.9. It is evident from these figures that the ME coefficient (dE/dH) exhibits maximum response for small field and then decreases with increasing DC magnetic field. Magnetostrictions in the magnetic phase get saturated at the higher value of applied magnetic field which produces a constant electric field in piezoelectric phase [53-54]. It is evident from Fig.9(f) that the value of ME coefficient was found to higher than 32 mV/cm-Oe for the ferroelectric rich composite composition with x=0.2-0.5. ME coefficient drastically decreases for the compositions with x=0.8 and 0.9 due to lower resistivity of samples in magnetic phase rich compositions. Similar trend of variations of ME coefficient is observed in $BaTiO_3$-$CoFe_2O_4$ [55], $Ni_{0.5}Co_{0.5}Fe_2O_4$/$Ba_{0.8}Pb_{0.2}TiO_3$ [56], $Ba_{0.8}Pb_{0.2}TiO_3$/$Cu_{0.4}Co_{0.6}Fe_2O_4$ [57], $Ni_{0.3}Cu_{0.4}Zn_{0.3}Fe_2O_4$/$BaTiO_3$-PZT [53] and $Ni_{0.9}Zn_{0.1}Fe_2O_4$/PZT [58-59] multiferroic particulate composites.

As discussed earlier, high temperature sintering of composites has more chances to taking reaction between constituent phases and formation of unwanted impurity phases which will affect the ME response. Reaction between constituent phases of composites can be controlled by lowering the sintering temperature or reducing the sintering time. An approach to lower the sintering temperature is the fabrication of composite by nano-powders of ferroelectric and magnetic phases where large surface to volume ratio will provide good mechanical contact and better ME response. Sintering time can be reduced by following the various method of synthesis like spark plasma, hot press and RF sputtering. ME coefficient has been obtained to be ~ 850 mV/cm-Oe in $BiFeO_3$/$Ni_{0.5}Zn_{0.5}Fe_2O_4$ composite prepared via RF-sputtering [19]. Significant ME response has been reported in $Ba(Zr_{0.2}Ti_{0.8})O_3$-$(Ba_{0.7}Ca_{0.3})TiO_3$/$CoFe_2O_4$ bulk composite meanwhile the sintering temperature is as high as ~ 1200 $^0$C [17,18]. Non-centrosymmetric phase transformation property in the ferroelectric phase also plays an important role in the enhancement of the ME property in the composite. Phase transformation which introduces the strain in the ferroelectric phase provides the better coupling between two phases in the composite. In the laminate composite of $La_{0.67}Sr_{0.33}MnO_3$ and $BaTiO_3$, substrate $BaTiO_3$ undergoes a phase transition from the tetragonal phase to monoclinic phase at the temperature 278 K. The structural transformation in substrate $BaTiO_3$ exhibits huge magnetization (70 %) where the strain changes by only 1 %. [2,34]. In (1-x)BNT-PT/xNZFO, spontaneous lattice strain is 2 % for the composition x=0 which changed to 11 % for x=0.8. Enhancement in the tetragonality in BNT-PT is linked with the partial diffusion of ions $Fe^{3+}$ ions at B-site of the ferroelectric phase as reported for $BiFeO_3$-$PbTiO_3$ solid solutions [36-38]. $BiFeO_3$-$PbTiO_3$ is a unique perovskite in which tetragonality decreases with increasing the $PbTiO_3$ content but increases with increasing the concentration $BiFeO_3$. Above discussion depicts that enhancement of tetragonality is linked with the diffusion of $Fe3^+$ ions in ferroelectric phase where the compositions exhibiting large tetragonality finally transform into rhombohedra phase ($R\bar{3}c$) to minimize the energy by lowering the strain of the system [38]. In addition to the reported literatures, evidence of partial $Fe^{3+}$ ions diffusion in the ferroelectric BNT-PT phase of our composite has been also found from EDS study. This phenomenon is completely opposite to the $PbZr_xTi_{1-x}O_3$ [60], $Pb(Mg_{1/3}Nb_{2/3})O_3$-$xPbTiO_3$ [61], $Pb(Sc_{1/2}Nb_{1/2})O_3$-$xPbTiO_3$ [62] and $Bi(N_{1/2}Ti_{1/2})O_3$-$PbTiO_3$ [22] solid solutions, where the tetragonality increases with increasing the $PbTiO_3$ concentration. In $BiFeO_3$-$PbTiO_3$ high tetragonality has been explained by covalent nature of 'A-O' and 'B-O' bonds which increases the tetragonality [36]. However, high tetragonality in the $Bi(Zn_{1/2}T_{1/2})O_3$-$PbTiO_3$ has been explained by strong coupling between A-site and B-site cations displacement using *ab initio* calculations, meanwhile tetragonal to rhombohedral structural distortion has not been reported in this system [63].

## 4. Conclusions

In summary, multiferroic $(1-x)0.51Bi(Ni_{1/2}Ti_{1/2})O_3$-$0.49PbTiO_3/xNi_{0.6}Zn_{0.4}Fe_2O_4$ based particulate composites have been fabricated by taking the magnetic phase NZFO and morphotropic phase boundary composition of ferroelectric phase BNT-PT aiming to get high ME response. The lattice parameters of both the magnetic and ferroelectric phases are modified evident the partial ionic diffusion between two phases. Tetragonality of ferroelectric phase BNT-PT continuously increases with increasing the NZFO content and finally turns into rhombohedral ($R\bar{3}c$) phase for x=0.9 by lowering the strain up to 0.7 % to minimize the energy of the system. The maximum ME coefficient for (1-x)BNT-PT/xNZFO is obtained in the ferroelectric rich compositions. To get the large ME-coefficient in composite which is mediated by strain, a composition close to MPB may be chosen for ferroelectric phase $Bi(M'_{1/2}M''_{1/2})O_3$-$PbTiO_3$, the structure of which latter may get modified to tetragonal phase. Since most of the $Bi(M'_{1/2}M''_{1/2})O_3$-$PbTiO_3$ type ferroelectrics are sintered at lower temperatures (~ 1000 $^0$C), it's composite formation will be free from unwanted secondary impurity phases which will provide good ME response. It is obvious that at high sintering temperatures, the magnetic and ferroelectric phases may react without diffusion of ions. We have found that the magnetic ordering may be tuned by subsequent composite formation which turns into superparamagnetic to ferrimagnetic ordering in (1-x)BNT-PT/xNZFO composite. This work will further encourage the authors to investigate new multiferroic particulate composites sintered at low temperature by taking various $Bi(M'_{1/2}M''_{1/2})O_3$-$PbTiO_3$ types ferroelectrics and $(Ni,Zn)Fe_2O_4$, $(Co,Zn)Fe_2O_4$, and $(La,Sr)MnO_3$ magnetic phases.

**Acknowledgment**

Authors are thankful to Mr. Neeraj Piplani, Marine-India for ME measurements.

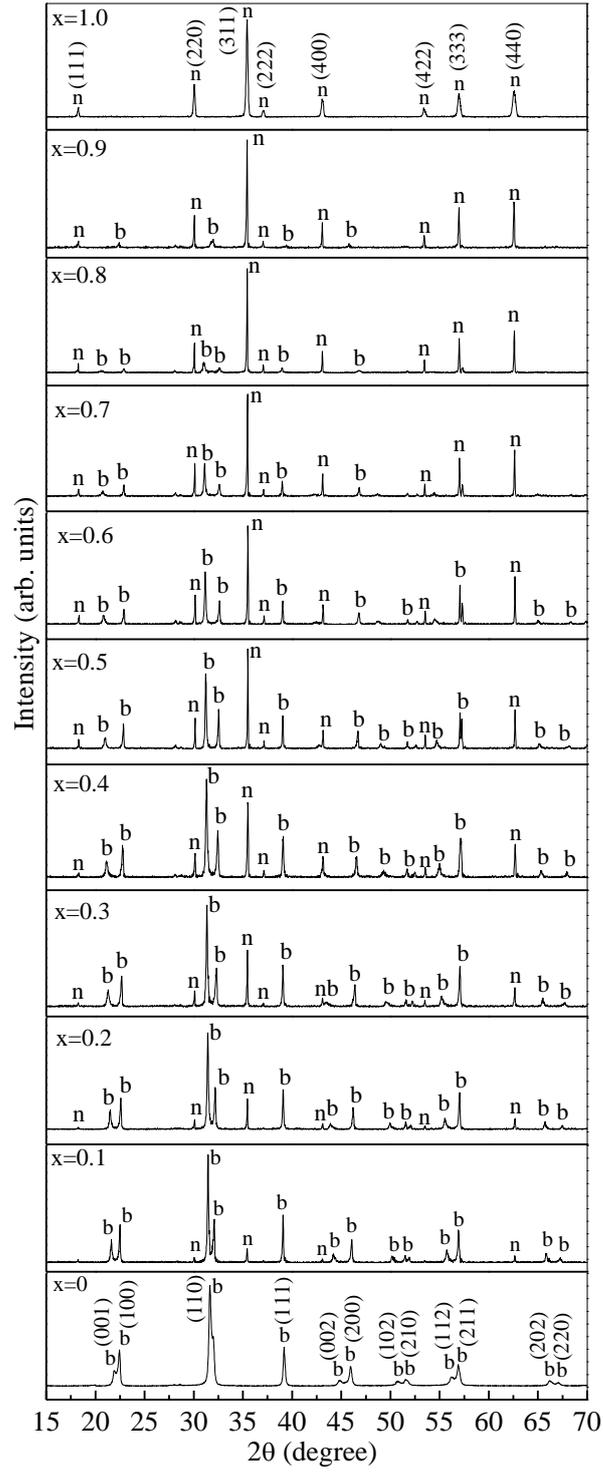

**Fig.1** Powder XRD profile of (1-x)BNT-PT/xNZFO composite sintered at 1000 $^{0}$C for the compositions with x=0, 0.1, 0.2, 0.3, 0.4, 0.5, 0.6, 0.7, 0.8, 0.9 and 1.0. Ferroelectric phase BNT-PT and magnetic phase NZFO are marked by 'b' and 'n', respectively.

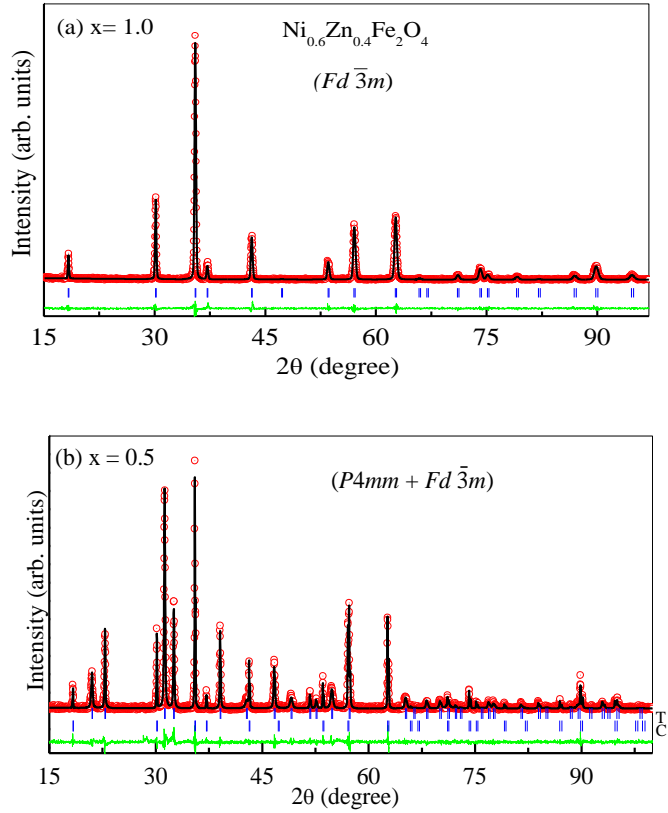

**Fig.2** Observed (dots), calculated (continuous line) and difference (continuous bottom line) Rietveld structural profile fit of (a) NZFO using cubic space group ($Fd\bar{3}m$) and, (b) BNT-PT/NZFO composite for the composition x=0.5 using cubic ($Fd\bar{3}m$) + tetragonal ($P4mm$) space groups. The vertical tick marks above the difference plot show the peak positions for cubic (C) and tetragonal (T) phases.

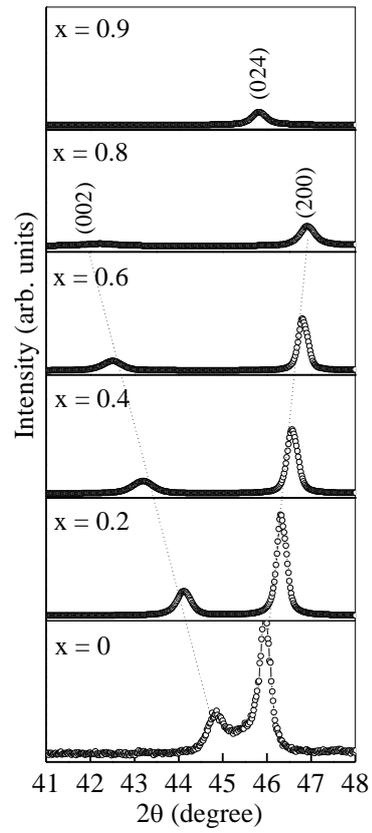

**Fig.3** Tetragonal (002) and (200) XRD peaks of BNT-PT for the compositions x=0.1-0.8 and rhombohedral peak (024) for the composition x=0.9.

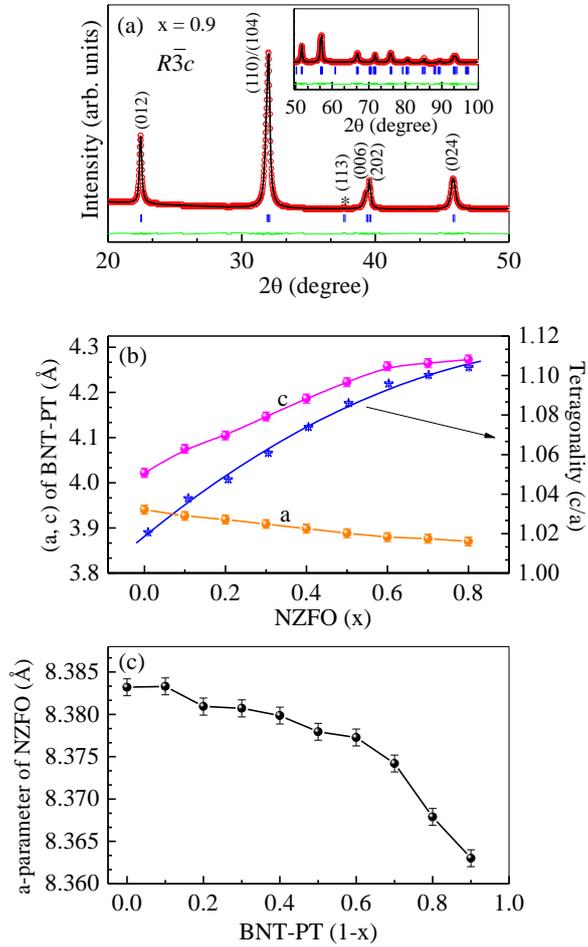

**Fig.4 (a)** Observed (dots), calculated (continuous line) and difference (continuous bottom line) Rietveld fit for the composition with x=0.9 using $R\bar{3}c$ space group. **(b)** Variation of lattice parameters and tetragonality of ferroelectric phase BNT-PT for the compositions x=0-0.8 **(c)** Variation of lattice parameter of magnetic phase NZFO for the compositions x=0.1-1.0.

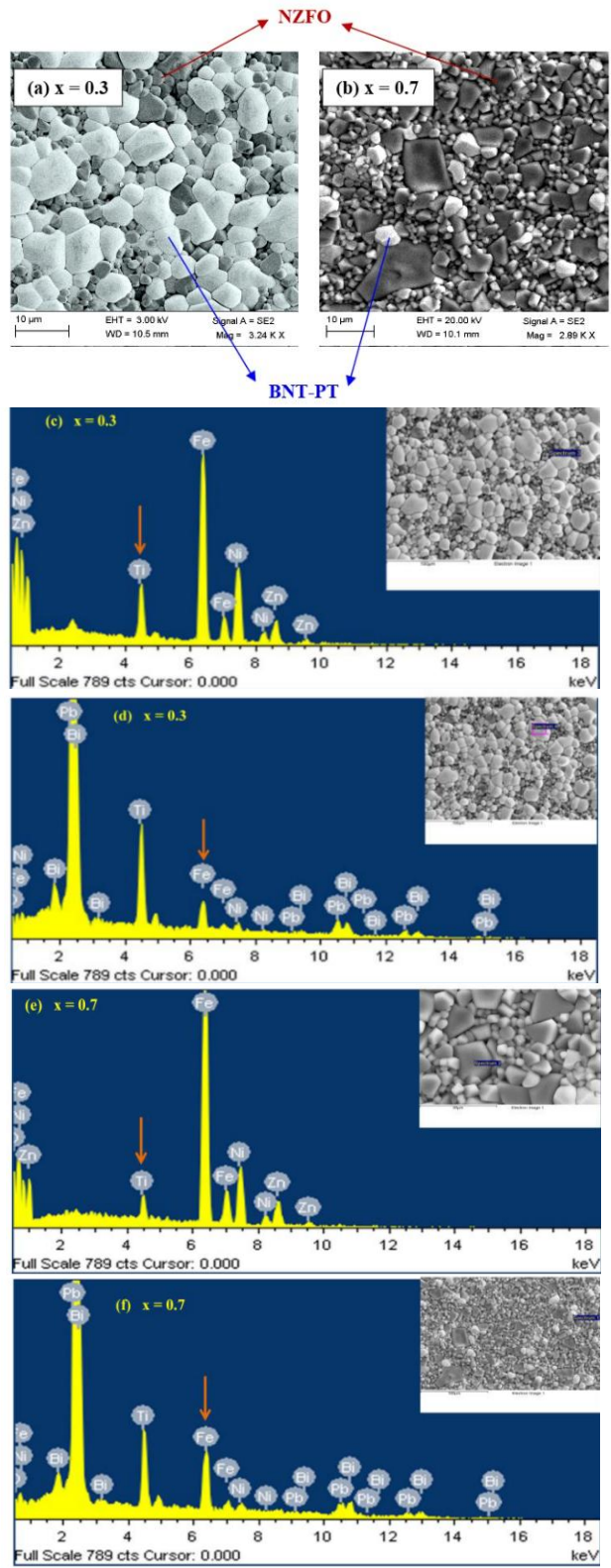

**Fig.5** SEM image for the compositions with **(a)** x=0.3 and **(b)** x=0.7 **(c)** EDS spectrum of the dark grains (NZFO) for the composition with x=0.3 **(d)** EDS spectrum of the white grains (BNT-PT) for the composition x=0.3 **(e)** EDS spectrum of the dark grains for the composition x=0.7 and **(f)** EDS spectrum of the white grains for the composition x=0.7.

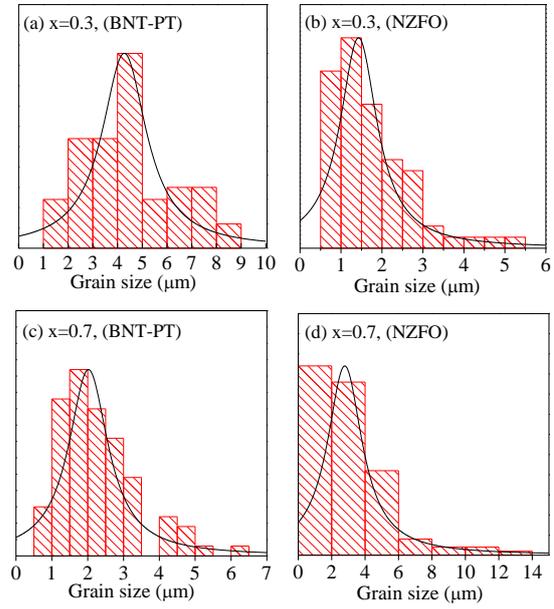

**Fig.6** Grain size distribution for **(a)** BNT-PT with x=0.3 **(b)** NZFO with x=0.3 **(c)** BNT-PT with x=0.7 and **(d)** NZFO with x=0.7.

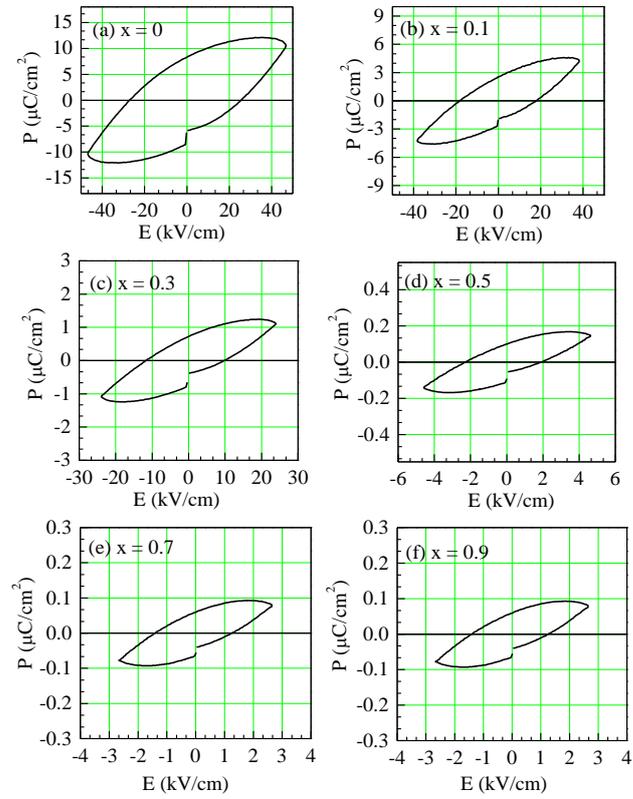

**Fig.7** (*P-E*) hysteresis loop for (1-x)BNT-PT/xNZFO composite for the compositions with **(a)** x=0 **(b)** x=0.1 **(c)** x=0.3 **(d)** x=0.5 **(e)** x=0.7 and **(f)** x=0.9.

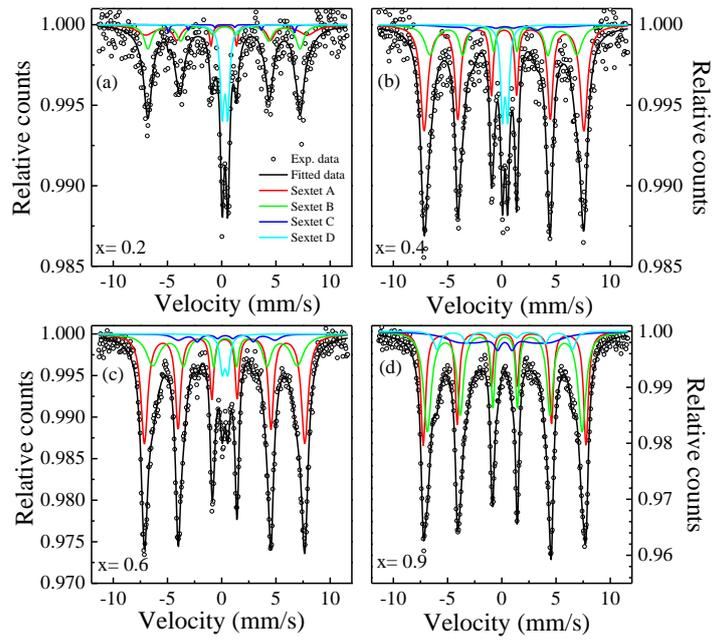

**Fig.8** Mössbauer spectra of (1-x)BNT-PT/xNZFO composite for the compositions with **(a)** x=0.2 **(b)** x=0.4 **(c)** x=0.6 and **(d)** x=0.9.

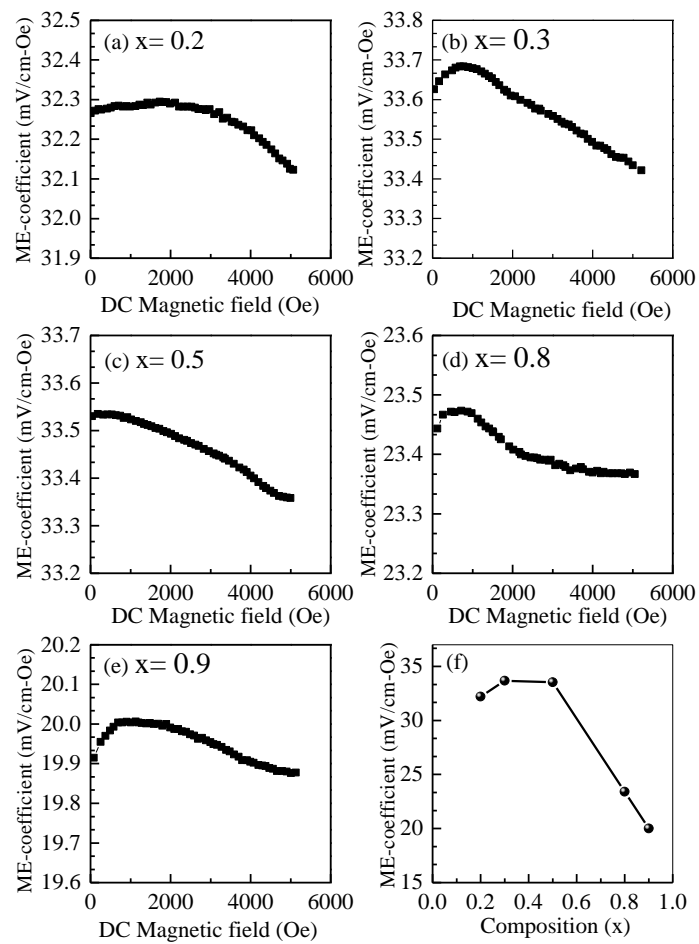

**Fig.9** Variation of linear ME-coefficient (dE/dH) with applied DC-magnetic field for (1-x)BNT-PT/xNZFO composites with compositions **(a)** x=0.2 **(b)** x=0.3 **(c)** x=0.5 **(d)** x=0.8 and **(e)** x=0.9. **(f)** Variation of linear ME-coefficient (dE/dH) with composition (x).

**Table 1:** Refined structural parameters and agreement factors for NZFO using cubic space group $Fd\bar{3}m$.

| Ions | x | y | z | B (Å$^2$) |
|---|---|---|---|---|
| $Zn^{2+}/Fe^{3+}_{I}$ | 0.125 | 0.125 | 0.125 | 0.5(1) |
| $Ni^{2+}/Fe^{3+}_{II}$ | 0.5 | 0.5 | 0.5 | 0.2(1) |
| $O^{2-}$ | 0.2567(3) | 0.2567(3) | 0.2567(3) | 0.6(2) |
| | $R_{wp}$=12.5, | $R_p$=8.5, | a=8.3823(3) Å, | $\chi^2$=1.24 |

**Table 2:** Refined structural parameters and agreement factors for BNT-PT/NZFO (x=0.5) composite taking cubic space group $Fd\bar{3}m$ and tetragonal space group (*P4mm*).

| | NZFO | | ($Fd\bar{3}m$) | |
|---|---|---|---|---|
| **Ions** | x | y | z | B (Å$^2$) |
| $Zn^{2+}/Fe^{3+}_{I}$ | 0.125 | 0.125 | 0.125 | 1.0(1) |
| $Ni^{2+}/Fe^{3+}_{II}$ | 0.5 | 0.5 | 0.5 | 1.1(1) |
| $O^{2-}$ | 0.2585(3) | 0.2585(3) | 0.2585(3) | 1.1(2) |
| | | a= 8.3779(1) Å | | |
| | **BNT-PT** | | (*P4mm*) | |
| $Bi^{3+}/Pb^{2+}$ | 0.0 | 0.0 | 0.0 | 3.0(1) |
| $Ni^{2+}/Ti^{4+}$ | 0.5 | 0.5 | 0.54(2) | 0.7(1) |
| $O_1$ | 0.5 | 0.5 | 0.16(1) | 1.0(0) |
| $O_2$ | 0.5 | 0 | 0.64(2) | 1.0(0) |
| $R_p$=9.4, | $R_{wp}$=14.6, | a= 3.8883(1) Å, | c= 4.2227(2) Å, | $\chi^2$=1.57 |

**Table 3:** The Hyperfine magnetic field ($H_{hf}$), isomer shift ($\delta$), quadrupole splitting ($\Delta$), line width ($\Gamma$) and Relative Area ($R_A$) in percentage of tetrahedral and octahedral sites of $Fe^{3+}$ ions for (1-x)BNPT/xNZFO composites derived from Mössbauer spectra recorded at room temperature. Isomer shift values are relative to Fe metal foil ($\delta = 0.0$ mm/s).

| x= | Iron Sites | Relative Area ($R_A$) % | Inner linewidth, ($\Gamma$) mm/s | Isomer shift, ($\delta$) mm/s | Quadrupole splitting, ($\Delta$) mm/s | Hyperfine field ($H_{hf}$) Tesla | Fitting quality ($\chi^2$) |
|---|---|---|---|---|---|---|---|
| 0.2 | Sextet A | 23.7 | 0.312 | 0.317 | 0.246 | 46.2 | 0.914309 |
| | Sextet B | 59.5 | 0.739 | 0.244 | -0.07 | 43.5 | |
| | Sextet C | 1.7 | 0.23 | 0.599 | 0.609 | 36.1 | |
| | Doublet | 15.1 | 0.426 | 0.305 | 0.511 | -- | |
| 0.4 | Sextet A | 57.5 | 0.404 | 0.202 | -0.001 | 45.6 | 0.914348 |
| | Sextet B | 25.0 | 0.437 | 0.236 | -0.173 | 42.1 | |
| | Sextet C | 11.4 | 0.941 | 0.381 | 0.228 | 31.3 | |
| | Doublet | 6.1 | 0.406 | 0.294 | 0.183 | -- | |
| 0.6 | Sextet A | 56.4 | 0.433 | 0.264 | 0.001 | 45.8 | 1.3867 |
| | Sextet B | 32.9 | 0.473 | 0.316 | -0.015 | 41.3 | |
| | Sextet C | 7.2 | 0.612 | 0.393 | 0.127 | 27.7 | |
| | Doublet | 3.6 | 0.506 | 0.309 | 0.512 | -- | |
| 0.9 | Sextet A | 26.5 | 0.368 | 0.249 | 0.025 | 46.5 | 1.54598 |
| | Sextet B | 39.2 | 0.383 | 0.290 | -0.032 | 44.2 | |
| | Sextet C | 10.7 | 0.706 | 0.325 | -0.109 | 38.7 | |
| | Sextet D | 23.6 | 0.724 | 0.253 | -0.043 | 26.2 | |